\documentstyle[aps]{revtex}

\begin{document}
\draft
\renewcommand{\thesection}{\Roman{section}}
\begin{center}

{\Large \bf Spatially Periodic Orbits in Coupled Sine Circle Maps}
 
\end{center}
 
\vspace{10pt}
 
\begin{center}

{Nandini Chatterjee$^1$\footnote{Electronic Mail : nandini@imsc.ernet.in} and Neelima Gupte$^2$ \footnote{Electronic Mail : gupte@imsc.ernet.in} }
 
\vspace{5pt}
 
{$^1$ Department of Physics,
 
University of Pune,
 
Pune 411007, INDIA.}
 
\vspace{7pt}
 
{$^2$ Department of Physics,
 
Indian Institute of Technology, Madras
 
Madras 600036, INDIA.}
 
\end{center}

\vspace{6pt}

\date{\today}

\vspace{10pt}

\begin{abstract}

We study spatially periodic orbits for a coupled map lattice of sine circle 
maps with nearest neighbour coupling and
periodic boundary conditions. 
The stability analysis for an arbitrary spatial period $k$  is carried
out in terms of the independent variables of the problem and 
the stability matrix is reduced to a  neat block diagonal form.
For a lattice of size $kN$, we show that the largest eigenvalue for the
stability matrix of size $kN \times kN$ is the same as that for the 
basic spatial period $k$ matrix  of size $k \times k$.
Thus  the analysis for a $kN$ lattice case can be
reduced to that for a $k$ lattice case. We illustrate this explicitly 
for a spatial period two case. Our formalism is general and can be extended to
any coupled map lattice.
We also obtain the stability regions of solutions which
have the same spatial and temporal period numerically. Our analysis
shows that such regions form a set of Arnold tongues in the
$\Omega-\epsilon-K$ space. 
The tongues corresponding to higher spatial periods are contained 
within the tongues seen in the temporally periodic spatial period 
one or synchronised case. 
We find an interesting  new bifurcation wherein the 
the spatially synchronised and
temporal period one solution undergoes a bifurcation to a
spatio-temporal period two travelling wave solution. The edges of the
stability interval of this solution are analytically obtained. 
\end{abstract}
\vspace{10pt}
PACS number(s) : 05.45. +b

\newpage

It is well known that systems with many degrees of freedom show
complex spatio-temporal behaviour \cite{crutch 87}. This behaviour may lead to 
highly
structured self-organising patterns, or may be highly incoherent in the
spatial or  temporal sense. Studies of systems like Josephson-Junction
lattice arrays \cite{weis 95,cher 95}, 
 multi-mode lasers \cite{rroy 94}, reaction-diffusion equations \cite{kap 93}, 
charge density waves \cite{sompo 91} and biological systems \cite{strogatz 89,gray 89} 
have provided evidence of the rich variety of behaviour that appears  
in such extended systems. The analysis of such systems via nonlinear 
partial differential
equations has proved to be extremely difficult. On the other hand, the 
modelling of spatially extended
systems by coupled, dynamically evolving elements situated at the discrete 
sites of a lattice 
 has turned out to be analytically and computationally tractable and has 
yet succeeded in capturing phenomena like
spatio-temporal intermittency \cite {ch 87},
pattern formation \cite {al 92}, phase separation \cite{op1 87,el 88,oo 87} and
chemical wave propagation processes \cite{kap 93,kap1 91} which occur in realistic systems.

A striking feature of the spatially extended systems described above, is the existence
of regimes where the systems show highly structured and spatially coherent
behaviour. Spatially periodic behaviour has been seen in coupled
oscillator arrays \cite{choi 94,op 86}, charge density waves
\cite{su 87}, biological systems \cite{gray 89,davis 85} and many others.
There has been much recent interest in the study of the simplest
spatially periodic mode, that of synchronized
behaviour in a variety of systems like
coupled oscillator arrays
\cite{wa 84,kwies 89,heagy 94}, coupled pendulums \cite {abraham 93},
electronic oscillator circuits \cite {ash 90}, and in pattern formation
\cite {kk 89}. Systems of coupled sine circle maps \cite{nc 96} appear to 
encapsulate some
of the generic behaviour found in such systems. Studies of spatially 
periodic behaviour in coupled sine circle map
lattices may yield some insight into the spatially periodic behaviour
seen in such systems.

We study the stability of spatially periodic behaviour in a one
dimensional system of coupled sine circle maps with 
nearest neighbour diffusive coupling and periodic
boundary conditions. The problem is recast in terms of convenient sum
and difference variables which reduce the stability matrix to a neat
block diagonal form. We set up the problem for a lattice of size $kN$,
where $k$ is the basic spatial period and $N$ is the number of copies of
this basic spatial period that can be contained in the lattice. The
stability matrix in this case has size $kN \times  kN$. We reduce this
matrix to a block diagonal form of $N$ blocks each of size $k \times k$.
Further, we obtain the general form for these $k \times k$ matrices,
and use the Gerschgorin theorem to show that the 
largest eigenvalue of the $kN \times kN$ matrix is the
same as that of the $k \times k$ matrix corresponding to the basic
spatial period. Thus the stability analysis for a lattice of size $kN$
is reduced to that for a lattice of size $k$. 

We carry out the numerical stability analysis of spatially
periodic solutions. In particular  
we find numerically the regions of stability in parameter space of 
spatially periodic solutions whose
temporal period is the same as their spatial period . We find that such 
solutions 
 form  a set of Arnold tongues in the three parameter $\Omega - \epsilon - K$
space of 
the coupled sine circle map system.
Solutions corresponding to higher spatial periods are contained within
the tongues seen in the temporally periodic spatial period one or
synchronised case \cite{nc 96}. We also obtain a Devil's staircase for
such solutions at $K=1$. An interesting  observation from the Arnold
tongue plot is the existence of
a new bifurcation within
the largest tongue, wherein the
the spatially synchronised and
temporal period one solution undergoes a bifurcation to a
spatio-temporal period two travelling wave solution. The stability
analysis of the travelling wave solution can be very conveniently
carried out in terms of the new variables and  
the edges of the stability interval of this solution are analytically obtained.

The paper is organized as follows.
In section I we consider a lattice of sine circle maps with nearest
neighbour diffusive coupling and periodic boundary conditions for
a lattice of $kN$ sites which can support  a basic spatial period $k$ which
is replicated $N$ times. (See Fig.1a ($k=2$ and
$N =3$) for an illustration of this ). We first
analyse the linear version or  the coupled shift map case 
and obtain the largest eigen value of the stability matrix for a solution with spatial period $k$, and then set up the stability matrix for the full
nonlinear version. 
We identify the independent variables for any arbitrary
spatial period $k$, and carry out a linear stability analysis for such
a solution for any arbitrary  temporal behaviour in terms of these
independent variables. 
In section II, we carry out 
similarity transformations using direct products of Fourier matrices
which reduce a $kN \times kN$ stability matrix to $N$ blocks each of size $k \times k$.
Further reduction of the eigen-value analysis of $N$ matrices of size $k \times k$ to that of a single matrix
of size $k \times k$, is achieved in section II.A.
We  demonstrate that the largest
eigenvalue for a spatial period $k$ solution in a lattice of
$kN$ sites is the same as the largest eigenvalue for a  spatial
period $k$ solution in a lattice of $k$ sites. As a result, it is sufficient
to look at the eigenvalues of the basic spatial period i.e. the
$k \times k$ matrix instead of a $kN \times kN$ matrix
to obtain the stability conditions for  
a spatial period $k$ solution. 
An explicit demonstration for a spatial period two case
which is special is carried out in section II.B.
Section III discusses numerical simulations for various spatiotemporally
periodic solutions. We obtain the regions of stability of various spatiotemporal
solutions with the same spatial and temporal period in the $\Omega - \epsilon - K $ space. 
We obtain interesting
plots which resemble the Arnold tongues seen for the spatially synchronized
 and temporally periodic solution \cite{nc 96} of the $\Omega - \epsilon - K $ space.
These tongues are contained within the tongues obtained for the spatially
synchronized solution  and also display  the symmetry about $\Omega = {1 \over 2}$. 
We obtain the Devil's  staircase for $K=1$ in the $\Omega - \epsilon - {P \over Q} $ 
space and discuss
its features. We obtain an interesting new bifurcation in the temporal 
period one tongue ($\Omega = 0$ and $\Omega = 1$)
 which gives rise to a spatial period two, temporal period two
solution which is of the travelling wave type.
In section IV  we discuss the travelling wave solution. We carry out a linear stability
analysis for the travelling wave solution in terms of  the independent variables
defined earlier 
which turn out to be  convenient for this analysis. We then obtain analytically
the edge for the $\epsilon$ interval for such a travelling wave solution
with spatial period two and temporal period two at $\Omega = 0$ and $K=1$.
We conclude with section V which discusses results and applications
to various physical situations. 

\section{The model and formulation}

The specific model under study, is a 1-dimensional coupled map lattice of
sine-circle maps with nearest neighbour diffusive symmetric normalized
coupling (also called
future coupled laplacian coupling) and periodic boundary conditions 
\cite{nc 96,bak 84}. This is defined by the evolution equation 

\begin{eqnarray}
\theta_{t+1}(i) & = & (1-\epsilon) \Big( \theta_{t}(i) + \Omega -
{K \over(2 \pi)} \sin(2 \pi\theta_{t}(i)) \Big) \nonumber \\
 & + &{\epsilon \over2}
\Big\{\theta_{t}(i+1)
  +  \Omega -{K \over(2 \pi)} \sin(2 \pi\theta_{t}(i+1)) \nonumber \\
  &+ &
\theta_{t}
(i-1) + \Omega
 - {K \over(2 \pi)} \sin(2 \pi\theta_{t}(i-1)) \Big\} \hspace{0.2 in} mod \hspace{0.05 in} 1
\end{eqnarray}
where $\theta_{t}(i)$ is the angular variable associated with the $i$th lattice site,
at time $t$ and lies between $0$ and $1$, $K$ is the strength of the nonlinearity, 
$\Omega$ is the period of the system for $K=0$ and $\epsilon$ which lies between 
0 and 1 is the strength of the coupling
parameter. We study the system for the homogeneous parameter case where $\Omega$ and $K$ take
the same value at each lattice site. However
the framework we develop can be used to study  inhomogenous systems  where the values
of  $\Omega$ and $K$  depend on the lattice site.

We consider a lattice of $kN$ sites and are interested in periodic
solutions such that 
$k$ is the basic spatial period 
and $N$ is the number of copies of this
basic spatial period. For example, in Fig.1(a), we have
a lattice of six lattice sites for which alternate lattice sites have
the same value. Hence it is a spatial period two solution and
$k=2$. This basic block is repeated three times so we have
$N=3$. In this paper we show how the
stability analysis for a lattice of
$kN$ sites can be reduced to just the study of the 
basic period of the $k$ lattice site case. 
We show that to study
the stability properties of a spatial period two
solution in a  lattice of six sites, instead of
looking at the eigen-value of the full $6 \times 6$ dimensional stability matrix  
it is sufficient
to look at the largest eigenvalue arising out of the period two solution
of a lattice of two sites i.e. the eigenvalue of  a matrix of size 
$2 \times 2$.  

\subsection{The shift map case}
We begin with the simplest case which is $K=0$, i.e. the coupled
shift map case. This is just the linear version of the circle map
and is a much simpler system to  study than the full non-linear version \cite{nc 96}.

 We consider a system of $kN$ coupled shift maps
with nearest neighbour diffusive symmetric coupling
and periodic boundary conditions. 
As illustrated in Fig.1(a) (where $k =2$ and $N=3$), this is a lattice
which can support a solution with basic spatial period $k$ and $N$ replica solutions. 
 The evolution equations are 
 \begin{equation}
 \theta_{t+1}(i) = (1- \epsilon)\Big( \theta_{t}(i) + \Omega \Big) +
 {\epsilon \over 2} \Big( \theta_{t}
 (i + 1) + \theta_{t}(i-1) \Big) + \epsilon \Omega
 \hspace{0.2 in} mod \hspace{0.05 in} 1
 \end{equation}

We observe that for a spatial period $k$ solution, at any time $t$, 
the value of the variable at the $i$th lattice site is the same
as the value at the $(i + k)$th site. Thus, the difference between the
variable values of the $i$th and the $(i + k)$th lattice site, 
approaches zero for all such pairs of neighbours.

Setting up the evolution equation for such a difference we have 
\begin{eqnarray}
  \theta_{t+1}(i) - \theta_{t+1}(i+k) & = & (1 - \epsilon) \Big( \theta_{t}(i)
   - \theta_{t}(i+k) \Big) \nonumber \\
   & + & {\epsilon \over 2} \Big( \theta_{t}(i+1) -
   \theta_{t}(i+k+1) + \theta_{t}(i-1) - \theta_{t}(i+k-1) \Big)
   \end{eqnarray}

Eq. 3 can be completely expressed in terms of the differences
$ a_{t}^{k}(i)$ defined as 
\begin{equation}
  a_{t}^{k}(i) = \theta_{t}(i) - \theta_{t}(i+k)
 \end{equation}
where the superscript $k$ denotes the spatial period at any time $t$.

The differences thus evolve as 
 \begin{equation}
   a_{t+1}^{k}(i) = (1 - \epsilon)(a_{t}^{k}(i)) + {\epsilon \over 2} \Big( a _{t}^{k}(i+1) 
  + a_{t}^{k}(i-1) \Big) \hspace{0.2 in} mod\hspace{0.05 in} 1
 \end{equation}
It can be easily seen that  $a_{t}^{k}(i) = 0$, is a spatial
period $k$ solution for Eq.5.

  Expanding upto the linear term about this solution leads to
  a stability matrix $J^{kN}_{t}$ \cite{FN}
 given by
 
\begin{equation}
   J^{kN}_{t} = \pmatrix{
(1- \epsilon)& {\epsilon \over 2} & 0 &0 & \cdots &  0 & 0& {\epsilon \over
2} \cr
{\epsilon \over 2} & (1- \epsilon)& {\epsilon \over 2} & 0 & 0 &\cdots & 0 &
 0\cr
0 &  {\epsilon \over 2} & (1- \epsilon)&  {\epsilon \over 2} & 0 & \cdots & 0
& 0\cr
   \vdots&\vdots&\vdots&\vdots&\vdots&\vdots&\vdots&\vdots \cr
   {\epsilon \over 2} & 0 & 0&\cdots &  0 & 0& {\epsilon \over 2} &
  (1- \epsilon)\cr}
\end{equation}
                     This is a $kN \times kN$ matrix, which is also
 circulant and whose eigen values maybe explicitly obtained
 analytically.
    The eigenvalues of $J^{kN}_{t}$ are given by 
\cite{nc 96,davis 79}
      \begin{eqnarray}
      \lambda_{r} & = & (1- \epsilon)
       + 
        {\epsilon \over 2} ( \omega_{r}
+ \omega_{r}^{-1})
 \end{eqnarray}
  where $\omega_ {r}$ is the $kN$th root of unity given by\\
   \begin{equation}
    \omega_{r} = \exp{\Big({2i \pi (r - 1)\over kN}\Big)}
     \end{equation}
      On simplifying, this can be written as \\
       \begin{eqnarray}
       \lambda_{r} = (1- \epsilon)
 + \epsilon  \cos ({2 \pi (r - 1) \over kN})
     && \hspace{0.04 in}r = 1,2, \ldots, kN
\end{eqnarray}

The 
stability condition for  spatially periodic (with period $k$) orbits 
of the coupled shift map is given by  $\lambda^{largest} \leq 1$, the
largest eigenvalue of Eq. 9.  
It can be easily seen that the largest eigen value is +1, indicating that the coupled shift map is
 marginally stable. This is true for all spatial periods $k$ , including
the spatially synchronised case, $k=1$ \cite{nc 96}.  
Thus if we start off with initial conditions that correspond to a
spatially periodic solution, we remain on them. The temporal period
depends on the value of $\Omega$ and 
 we obtain temporally periodic orbits of period $Q$ for
rational values of $\Omega=P/Q$ and quasiperiodic orbits for
irrational values of $\Omega$.

\subsection{The coupled circle map case}
A  similar analysis can be carried out for a lattice of sine circle maps of $kN$ sites as defined
in Eq. 1. 
We look for the regions of stability
of spatially periodic solutions with spatial period
$k$  for a lattice of $kN$ sites. 

As in the coupled shift map case, here too we observe
that at time $t$ for a fixed $\Omega$ and $K$ and
a particular spatial period $k$
the difference between the $i$th and $(i+k)$th lattice sites
is zero. (Fig. 1a). The difference is again defined as
 \begin{equation}
 a_{t}^{k}(i)= \theta_{t}(i) - \theta_{t}(i+k)
\end{equation}

Using Eq.1 and setting up the equation of evolution for the differences
it can be easily seen that the evolution equation for these differences involves 
not just terms which involve the differences $a_{t}^{k}(i)$ but also
 terms of the kind
$(\theta_{t}(i) + \theta_{t}(i+k))$ which is just the sum of the
variables of the $i$th and $i+k$th site. 
 We also observe that at a fixed $\Omega$ and
$K$ and spatial period $k$ that 
the sum of
the $i$th and $i+k$th site is also a constant. 
So we now define
\begin{equation}
 b_{t}^{k}(i)= \theta_{t}(i) + \theta_{t}(i+k)
 \end{equation}
\hspace{4.5 in} $ \forall \hspace{0.1 in}i ; 1, \ldots kN $

Using Eq.1 we obtain the equations of evolution for $a_{t}^{k}(i)$ and $b_{t}^{k}(i)$ 
as 
\begin{eqnarray}
a_{t+1}^{k}(i)& =& (1-\epsilon)\Big(a_{t}^{k}(i) - {K \over \pi}
\sin( \pi a_{t}^{k}(i)) \cos( \pi b_{t}^{k}(i)) \Big) \nonumber \\
&+& {\epsilon \over2}\Big(a_{t}^{k}(i+1) - {K \over \pi}
\sin( \pi a_{t}^{k}(i+1)) \cos( \pi b_{t}^{k}(i+1)) \Big) \nonumber \\
&+& {\epsilon \over2}\Big(a_{t}^{k}(i-1) - {K \over \pi}
\sin( \pi a_{t}^{k}(i-1)) \cos( \pi b_{t}^{k}(i-1)) \Big) \nonumber
\hspace{0.2 in} mod\hspace{0.05 in} 1\\
\end{eqnarray}
and
\begin{eqnarray}
b_{t+1}^{k}(i)& =& (1-\epsilon)\Big(b_{t}(i) - {K \over \pi}
\sin( \pi b_{t}^{k}(i)) \cos( \pi a_{t}^{k}(i)) \Big) \nonumber \\
&+& {\epsilon \over2}\Big(b_{t}^{k}(i+1) - {K \over \pi}
\sin( \pi b_{t}^{k}(i+1)) \cos( \pi a_{t}^{k}(i+1)) \Big) \nonumber \\
&+& {\epsilon \over2}\Big(b_{t}^{k}(i-1) - {K \over \pi}
\sin( \pi b_{t}^{k}(i-1)) \cos( \pi a_{t}^{k}(i-1)) \Big) \nonumber  +
2 \Omega  \hspace{0.1 in} mod\hspace{0.05 in} 1\\
\end{eqnarray}

 It can be easily shown that,
 $\forall {i}$, $a_{t}^{k}(i) = 0$ and
    $ b_{t}^{k}(i) = s_{m} $ ($m \equiv i$ mod $k$), where $s_{1}, s_{2}, \ldots s_{k}$ are all
    distinct $ constants$,
    are solutions of Eq.12 and 13 for a fixed $\Omega$ and $K$.\\

To study the stability of any spatially periodic solution with
spatial period $k$  we need to examine
the eigenvalues of the linear stability matrix.
We expand Eqs. 12 and 13 about $ a_{t}^{k}(i) = 0$ and
  $ b_{t}^{k}(i) = s_{m}, m : 1,2 \ldots k $ distinct constants, upto the linear order
and obtain the matrix of coefficients $J_{t}^{2kN}$. (Since we have
two sets of equations namely for the $a_{t}^{k}(i)$'s and $b_{t}^{k}(i)$'s
each for $kN$ sites, the dimension of the matrix $J^{2kN}_{t}$ is
$2kN \times 2kN$). The largest
eigenvalue of this matrix crossing 1 determines the edge
of stability for the corresponding spatiotemporal solution. We have

\begin{equation}
J^{2kN}_{t} =  \pmatrix{ A ^{\prime {kN}}_{t} & B ^{\prime {kN}}_{t}\cr
   B ^{\prime {kN}}_ {t} & A ^{\prime {kN}}_{t}\cr}
   \end{equation}
where
\begin{equation}
 A ^{\prime {kN}}_{t} = \pmatrix{ \epsilon_{s}A_{t}^{k}(1) &
 \epsilon_{n} A_{t}^{k}(2) & 0 &
 \cdots & 0 & \epsilon_{n} A_{t}^{k}(kN) \cr
{\epsilon_n}A_{t}^{k}(1) &
 {\epsilon_s}A_{t}^{k}(2)  &
{\epsilon_n}A_{t}^{k}(3) &
    0  & \cdots & 0 \cr
    0 &
  \epsilon_{n}A_{t}^{k}(2) &
  \epsilon_{s}A_{t}^{k}(3) &
\cdots&  0 & 0 \cr
 \vdots& \vdots& \vdots& \vdots& \vdots& \vdots  \cr
      \epsilon_{n}A_{t}^{k}(1) & 0 &
         \cdots & 0 &
      \epsilon_{n}A_{t}^{k}(kN-1) &
  {\epsilon_s}A_{t}^{k}(kN) \cr}
 \end{equation}
 and
\begin{equation}
 B ^{\prime {kN}}_{t} =  \pmatrix{ \epsilon_{s}B_{t}^{k}(1) &
 \epsilon_n B_{t}^{k}(2) & 0 &
 \cdots &
 0 & {\epsilon_n}B_{t}^{k}(kN) \cr
 {\epsilon_n}B_{t}^{k}(1) &
 {\epsilon_s}B_{t}^{k}(2) &
 {\epsilon_n}B_{t}^{k}(3) & 0 &
   \cdots & 0 \cr
 0 & {\epsilon_n}B_{t}^{k}(2)
 & {\epsilon_s}B_{t}^{k}(3)
  & \cdots &
      0 & 0 \cr
\vdots& \vdots& \vdots& \vdots& \vdots& \vdots \cr
  {\epsilon_n}B_{t}^{k}(1) & 0 & \cdots & 0 &
   {\epsilon_n}B_{t}^{k}(kN-1) &
   {\epsilon_s}B_{t}^{k}(kN) \cr}
\end{equation}
      where
     $\epsilon_s , \epsilon_{n}$ are given by 

  \begin{equation}
 \epsilon_s = (1- \epsilon), \hspace{0.2in}
       \epsilon_n = {\epsilon \over 2}
       \end{equation}

Here, each
 \begin{eqnarray}
 A_{t}^{k}(i) = {\Big (} 1 - {K }\cos( \pi a_{t}^{k}(i))
  \cos( \pi b_{t}^{k}(i)) {\Big )}
  \end{eqnarray}
  and
 \begin{eqnarray}
 B_{t}^{k}(i) = {\Big (}  {K }\sin( \pi a_{t}^{k}(i))
   \sin( \pi b_{t}^{k}(i)){\Big )}
\end{eqnarray}
where $b_{t}^{k}(i)$ repeats after $k$ sites.
 
Imposing the conditions $a_{t}^{k}(i) = 0$ and $ b_{t}^{k}(i) = s_{m}$ where 
$  m:1,2 \ldots k $,
the stability matrix $J^{2kN}_{t}$ given by Eq.14 reduces to
a block diagonal form
\begin{equation}
J^{2kN}_{t} = \pmatrix{ M_{t}^{kN}& 0 \cr
    0& M_{t}^{kN}\cr}
        \end{equation}
        where the blocks $M^{kN}_{t}$ are identical
and each $M^{kN}_{t}$ is of the form,
  \begin{equation}
 M^{kN}_{t} = \pmatrix{ \epsilon_{s}\bar A_{t}^{k}(1) &
 \epsilon_{n} \bar A_{t}^{k}(2) & 0 & 
 \cdots & 0 & \epsilon_{n} \bar A_{t}^{k}(k) \cr
  \epsilon_{n}\bar A_{t}^{k}(1) &
 \epsilon_{s} \bar A_{t}^{k}(2) &
 \epsilon_{n}\bar A_{t}^{k}(3) &
0 & \cdots & 0 \cr
0 & \epsilon_{n} \bar A_{t}^{k}(2) & 
\epsilon_{s} \bar A_{t}^{k}(3) & 
\epsilon_{n} \bar A_{t}^{k}(4) & 0 &
 \cdots \cr
  \vdots& \vdots& \vdots& \vdots & \vdots & \vdots \cr
\epsilon_{n}\bar A_{t}^{k}(1) & 0 & \cdots & 0 &
\epsilon_{n}\bar A_{t}^{k}(k-1) & 
 \epsilon_{s} \bar A_{t}^{k}(k)\cr}
    \end{equation}
  and each
  \begin{equation}
    \bar  A^{k}_{t}(i) = {\Big (} 1
        - {K }
                 \cos( \pi s_{i} ) {\Big )}
                 \end{equation}
where, each $ m $ goes from \hspace{0.01 in} $ 1, \ldots, k $ and
repeats after $k$ sites and $m \equiv i mod k$.
We thus obtain the stability matrix in a block diagonal form
as given by Eq. 20 where each $M_{t}^{kN}$ is a matrix of size
$kN \times kN$.
We now carry out matrix transformations
which reduce the matrix $M_{t}^{kN}$ which is a $kN \times kN$
matrix into a block diagonal form with $N$ matrices each of size $k \times k$. Thus a $kN \times kN$
matrix is reduced to $N$ matrices each of size $k \times k$. Further we
use Gerschgorin's theorem for eigen values of square matrices
which reduces the stability analysis of the problem from the analysis of $N$ matrices of 
size $k \times k$
to that of just one matrix of size $k \times k$ - the basic spatial
period $k$ matrix.

\section{Reduction of the Stability Matrix }
 
We now illustrate the reduction of the matrix
$M_{t}^{kN}$ for a spatial period $k > 2$. The spatial period
 corresponding to $k=1$ is the synchronized
solution where $M_{t}^{kN}$ turns out to be  a circulant matrix
and  has been discussed in Ref.\cite{nc 96}. The case  $k=2$ has  special
features which will be  
 discussed in a separate section below.
We  consider the matrix $M_{t}^{kN}$ of Eq. 21 where $k > 2$.
 Using a similarity tranformation we now reduce the
matrix $M^{kN}_{t}$ which has dimensions $kN \times kN$ into 
a block diagonal matrix of $N$
blocks each of size $k \times k$.
 The similarity transformation
which achieves this, is a direct product of Fourier matrices of size $N \times N$ and
$k \times k$.

A Fourier matrix $F_{k}$ (where $k$ denotes the dimension of the matrix)
is defined as 
\begin{equation}
F^{*}_{k}(i,j)={1 \over \sqrt{k}}(\omega_{k}^{(i-1)(j-1)})
\end{equation}
$F_{k}$ is easily calculated as the conjugate transpose
of $F^{*}_{k}$ and $\omega_{k} = e^{2 \pi i \over k}$.

 The similarity
transformation for a  stability matrix of size  $kN \times kN$, which achieves the block
diagonal form  is given by
\begin{equation}
F_{kN}= F_{N} \otimes F_{k}
\end{equation}
where $\otimes$ indicates direct product.

For example, for a six lattice site case and a spatial period two solution,
we have $k=2$ and $N=3$ , i.e we have three replicas of the spatial
period two solution. Constructing Fourier matrices $F_{2}$ and
$F_3$ we obtain,
 
\begin{equation}
F^{*}_{2} = {1 \over {\sqrt 2}} \pmatrix{ 1 & 1 \cr
1 & -1 \cr}
\end{equation}
and 
\begin{equation}
F^{*}_{3} = {1 \over {\sqrt 3}} \pmatrix{ 1 & 1 & 1\cr
1 & {-1+ i \sqrt{3} \over {2}} & {-1- i \sqrt{3} \over {2}} \cr
1 & {-1- i \sqrt{3} \over {2}} & {-1+ i \sqrt{3} \over {2}} \cr}
\end{equation}

A direct product of $F_{3}$ and $F_{2}$ will give us a $6 \times 6$ matrix
which is required for the similarity transformation for a lattice of 6 sites where
the basic spatial period ($k$) is 2 and the number of copies ($N$) is 3.
Since Fourier matrices are also unitary the conjugate transpose of this matrix
is also its inverse and so $F^{*}_{k}$ is the inverse
of $F_{k}$.

The operation  $F_{kN}M_{t}^{k}F^{*}_{kN}$ gives the following block diagonal form for
$M_{t}^{kN}$ 

\begin{equation}
M_{t}^{kN} = \pmatrix{ M_{t}^{k}(1) & 0 & 0 & \cdots & 0 \cr
 0 & M_{t}^{k}(2) & 0 & 0 & \cdots \cr
\vdots & \vdots & \vdots & \vdots & \vdots \cr
0 & 0 & 0 & M_{t}^{k}(l-1) & 0 \cr
0 & 0 & 0 & 0 & M_{t}^{k}(l) \cr}
\end{equation}
where $l$ goes from $ 1, 2 \ldots N$.
 
For $k > 2$, each $M_{t}^{k}(l)$ is a $k \times k$ matrix and has a structure of the form

\begin{equation}
 M_{t}^{k}(l) = \pmatrix{ \epsilon_{s}\bar A_{t}^{k}(1) &
 \epsilon_{n}\bar  A_{t}^{k}(2) & 0 &
 \cdots & 0 & \epsilon_{n} \bar A_{t}^{k}(k) \omega_{l} \cr
{\epsilon_n}\bar A_{t}^{k}(1) &
 {\epsilon_s}\bar A_{t}^{k}(2)  &
{\epsilon_n}\bar A_{t}^{k}(3) &
    0  & \cdots & 0 \cr
    0 &  
  \epsilon_{n}\bar A_{t}^{k}(2) &
  \epsilon_{s}\bar A_{t}^{k}(3) &
 \epsilon_{n}\bar A_{t}^{k}(4) &  0 & \cdots \cr
 \vdots& \vdots& \vdots& \vdots& \vdots& \vdots  \cr
      \epsilon_{n}\bar A_{t}^{k}(1) \omega_{l}^{-1} & 0 &
         \cdots & 0 &
      \epsilon_{n}\bar A_{t}^{k}(k-1) &
  {\epsilon_s}\bar A_{t}^{k}(k) \cr}
\label{kay}
 \end{equation}
 where $\omega_ {l}$ is defined as  $\omega_l = e^{2 \pi i \over l}$ and
$\bar A_{t}^{k}(i)$ is as defined in Eq.22. A similar block diagonal
form will be achieved for the $k=2$ case as well but the matrix 
  $M_{t}^{k}(l)$ has a different
form for $k=2$ which  will be discussed later.

\subsection{Reduction to the basic spatial period - the $k \times k$ case $k > 2 $}

We now use Gerschgorin's theorem \cite{stein 67} for square matrices which
states that the eigen values of a matrix 
{$ A_m $} ( of dimension $m$ )  lie in
the union of all points $\Lambda$ such that
\begin{equation}
\left| \Lambda - a_{ii} \right| \leq C_{i}, i= 1,2 \ldots m 
\end{equation}
where
\begin{equation}
C_{i} = \sum_{j}^{\prime} |a_{ij}| \nonumber \\
\end{equation}

where $\sum_{j}^{\prime}$ indicates that the summation is for
$j= 1,2 \ldots, m$ excluding $j=i$ and $a_{ij}$ is the $ij$th
element of the matrix $A_{m}$.

By this theorem if one constructs circles in the complex plane
 for $ i = 1,2 \ldots m $ with the elements $a_{ii}$ as
centres and the sum of the absolute values of the remaining
elements in the $i$th row as the corresponding radii, then
each eigen value of $A_{m}$ must lie in or on one
of the circles, the Gerschgorin discs \cite{stein 67}.

We now consider a lattice of $k$ sites and the stability of
a spatial period $k$ solution. We have only one block for
a lattice of $k$ sites, and a spatial period $k$ solution, so
$l=1$ in Eq. 27 and $M_{t}^{k}(l)$ is given by
\begin{equation}
 M_{t}^{k}(1) = \pmatrix{ (1-\epsilon)c_{1} &
 {\epsilon \over {2}} c_{2} & 0 &
 \cdots & 0 & {\epsilon \over 2}c_{k} \cr
{\epsilon \over 2}c_{1} &
 {(1- \epsilon)}c_{2}  &
{\epsilon \over 2}c_{3} &
    0  & \cdots & 0 \cr
    0 &
  {\epsilon \over 2}c_{2} &
  (1- \epsilon)c_{3} &
  {\epsilon \over 2}c_{4} &  0 & \cdots \cr
 \vdots& \vdots& \vdots& \vdots& \vdots& \vdots  \cr
      {\epsilon \over 2}c_{1} & 0 &
         \cdots & 0 &
     {\epsilon \over 2}c_{k-1} &
  {(1- \epsilon)}c_{k} \cr}
 \end{equation}

where $c_{i}=(1-K\cos({\pi} b_{t}^{k}(i)))$ and $\omega_l=1$.

The Gerschgorin discs for this matrix will be $k$ circles of the 
following form.
\begin{eqnarray}
(x - (1-\epsilon)c_{i})^{2} + y^{2} & = & (|{\epsilon \over{2}}c_{i+1}| + |{\epsilon \over {2}}c_{i-1} |)^{2} 
\end{eqnarray}
where $i = 1 \ldots k$.

Thus according to Gerschgorin's theorem all the 
characteristic or eigen values of this matrix will lie
in or on these circles.

For a lattice of size $kN$, we now construct circles for the 
matrix $M_{t}^{k}(l)$ given
by Eq. 28. The general form of matrix $M_{t}^{k}(l)$ (Eq. 28)
 differs from that in Eq. 31
in the form of the first row and the last row. 
So while considering the
eigen values of $M_{t}^{k}(l)$ we need to construct Gerschgorin discs
only for the first row and the last row, the remaining
$k-2$ Gerschgorin discs will be the same as those obtained for the
$k$ lattice case above. 
For the $M_{t}^{k}(l)$ matrix we have the following
equation for the first Gerschgorin disc
\begin{eqnarray} 
(x_{1}-(1-\epsilon)c_{1})^2 + y_{1}^{2}  & = & ( | {\epsilon \over{2}}c_{2}| +
|{\epsilon \over{2}}c_{k}\omega_{l} |)^{2} \nonumber \\
& = & ( | {\epsilon \over{2}}c_{2}| +
|{\epsilon \over{2}}c_{k}| |\omega_{l} |)^2 
\end{eqnarray} 

Similarly the equation of the $k$th Gerschgorin disc is given by
\begin{eqnarray} 
(x_{k}-(1-\epsilon)c_{k})^2 + y_{k}^{2} & = & ( | {\epsilon \over{2}}c_{k-1}| +
|{\epsilon \over{2}}c_{1}\omega_{l}^{-1} | )^{2} \nonumber \\
& = & (|{\epsilon \over{2}}c_{k-1}| + | {\epsilon \over{2}}c_{1}|
 | \omega_{l}^{-1} |)^2  
\end{eqnarray} 

\vspace{0.2 in}

$\omega_{l}$ is defined as  $\omega_l = e^{2 \pi i \over l}$.
Hence $\left | \omega_{l} \right |$ and $\left | \omega_{l}^{l-1} \right|$ is 1.
Thus Eqs.33 and 34 reduce to 
\begin{equation}  
(x_{1}-(1-\epsilon)c_{1})^2 + y_{1}^{2} = ( | {\epsilon \over{2}}c_{2} | +
|{\epsilon \over{2}}c_{k} |)^{2} 
\end{equation}  
and
\begin{equation} 
(x_{k}-(1-\epsilon)c_{k})^2 + y_{1}^{2} = ( | {\epsilon \over{2}}c_{k-1}| +
|{\epsilon \over{2}}c_{1} |)^{2} 
\end{equation}   

So all the eigen values of the matrix $M_{t}^{k}(l)$
will lie in or on one of these circles. Thus the eigen values
of all the $N$ matrices (as $l$ : $ 1,2 \ldots N$) will lie
in or on one of these Gerschgorin circles.

But these are precisely the Gerschgorin discs we obtained 
for the matrix $M_{t}^{k}$ i.e. the matrix for a lattice
of $k$ sites and spatial period $k$ as obtained Eq. 32.

The stability of any spatially periodic solution is obtained by 
checking where the largest
eigenvalue crosses the unit circle, with $\lambda ^{largest}$ = $\pm 1$
being the condition for marginal stability.  
The analysis presented above shows that the bounds on the eigenvalues for
the $kN \times kN$ 
matrix given by Eqs. 32, 35 and 36 for the
$kN$ lattice case are the same as the bounds 
on the eigenvalues for the $k \times k $ matrix given by Eqs. 32
i.e.  that of the $k$ lattice case which is also the basic
spatial period. Thus the largest eigenvalue for $M_{t}^{k}(1)$ crosses 1
for the same parameter values as the largest eigenvalue of $M_{t}^{k}(l)$ ($l$ : $1, \ldots N$ ) crosses 1.
We thus conclude
that it is sufficient to check for  the largest eigenvalue of
the basic $k \times k$ matrix as given by
Eq.32 to decide the stability properties for any multiple of $k$ too
namely $kN$.
The bounds on the eigenvalue for a spatial period $k$ solution
for a lattice of $k$ sites will remain the same for a spatial period $k$ solution and a lattice of $kN$ sites.
We now illustrate this explicitly for a spatial period 2 case and thus a $2N$ case.

\subsection{The $k=2$ case - Illustration  for a $2N$ site lattice}

Consider the spatial period two case for two lattice sites which evolve via the evolution equations
12 and 13. (See Fig.1(a)).

Here $ \theta_{t}(i+2)$ is the same as $\theta_{t}(i)$

The stability matrix for a spatial period two solution is given by
\begin{equation}
M^{2}_{t}(1) = \pmatrix{ (1 - \epsilon)c_{1} & {\epsilon }c_{2} \cr
{\epsilon}c_{1} & (1- \epsilon)c_{2} \cr}
\label{two}
\end{equation}
Here
\begin{eqnarray*}
 c_{1} & = & (1-K\cos(\pi s_{1}))  \\
 c_{2} & = & (1-K\cos(\pi s_{2}))  
\end{eqnarray*}

It is clear from Eq. \ref{two} that the stability matrix for a spatial
period $k=2$, two lattice site case is not of the same form as that for $k > 2$. Since each of the two sites has just one
neighbour, each site is coupled to it's neighbour with a coupling
constant of strength $\epsilon$, and is coupled to itself with strength
$ 1- \epsilon$, in contrast with the $k > 2$ case where each site has
two neighbours and is
coupled to each of it's neighbours with a coupling constant of strength
${\epsilon \over 2}$ and to itself with strength $1- \epsilon$. The overall normalisation of the 
coupling in preserved in both cases. 
This results 
in the general form of the matrix $M_{t}^{2N}(l)$ having a different
form for the $k=2$ case as compared to the general $k$ case, $k > 2 $
(Eq. \ref{kay}).

Since $M_{t}^{2}$ is a $2 \times 2$ matrix the eigenvalues maybe be easily obtained
and are given by

\begin{equation}
\lambda = {(1-\epsilon)(c_{1}+c_{2}) \over{2}} \pm {1 \over 2} \sqrt{{(1 - \epsilon)}^{2}(c_{1}+c_{2})^{2} - 4(1- 2\epsilon)c_{1}c_{2}}
\end{equation}

It can be easily shown that the eigenvalue corresponding to the + sign
is the larger one.

 Now consider a lattice of $2N$ sites where $N$ is any positive integer. We carry out the above analysis for a spatial period two solution and obtain the matrix
$M_{t}^{2}(l)$ where $l$ : $ 1 \ldots N$. Any $M_{t}^{2}(l)$ is of the form

\begin{equation}
M^{2}_{t}(l) = \pmatrix{ (1 - \epsilon)c_{1} & {\epsilon \over2}(1+\omega_{l})c_{2} \cr
{\epsilon \over2}(1+\omega_{l}^{-1})c_{1} & (1- \epsilon)c_{2} \cr}
\end{equation} 
 
where $c_{1}$ and $c_{2}$ are as defined above and  $\omega_{l}$ is the $N$th root of unity and depends on the number
of replicas. For $N=3$ i.e. six lattice sites, $l : 1,\ldots 3$
 (in this case $k=2$). $ M^{2}_{t}(l)$ has a different form for the
$k=2$ case  as mentioned above.

The eigen-values of the  $2N$ case 
turn out to be
\begin{equation} 
\lambda = {(1-\epsilon)(c_{1}+c_{2}) \over{2}} \pm {1 \over 2} \sqrt{{(1 - \epsilon)}^{2}(c_{1}+c_{2})^{2} - 4c_{1}c_{2}(1- \epsilon)^{2} + c_1c_2 {\epsilon}^{2} {(2+ 2\cos({2{\pi}{(l-1) \over 2}}})) }
\end{equation}
 
The largest eigen value is obtained for  
$\cos({2 \pi (l-1) \over 2}) $ = 1 (which occurs for l = 1,3,5 \ldots). The
largest eigenvalue is  
\begin{equation}
\lambda = {(1-\epsilon)(c_{1}+c_{2}) \over{2}} + {1 \over 2} \sqrt{{(1 - \epsilon)}^{2}(c_{1}+c_{2})^{2} - 4(1- 2\epsilon)c_{1}c_{2}}
\end{equation}
Comparing the largest eigenvalue of 
Eq. 38 and that of Eq. 41 we find that the largest eigen value obtained for 
the $2N$ lattice case is the same as that obtained for the 2 lattice case. Thus 
for a $2N$
lattice it is sufficient
to look at the stability properties for just two lattice sites which is the
basic $k$ period. ($k$ = 2 in this case).

Gerschgorin's theorem  may also be used to prove the above. 
Using Gerschgorin's theorem  for the matrix defined
by Eq. 37 we find that all the eigen values of this matrix will
lie in or on the circles given by the equations
 
\begin{eqnarray}
 (x_{1}-(1-\epsilon)c_{1})^{2} + y_{1}^{2} = (\epsilon c_{2})^2 \nonumber \\ 
 (x_{2}-(1-\epsilon)c_{2})^2 + y_{2}^{2} = (\epsilon c_{1}) ^2
\end{eqnarray}
 
Constructing Gerschgorin discs again, now for Eq. 39 we have the following circles.
\begin{eqnarray}
 (x_{1}-(1-\epsilon)c_{1})^{2} + y_{1}^{2} & = &\left| {\epsilon \over {2}} c_{2}(1+\omega_{l}) \right|^{2} \nonumber \\
 (x_{2}-(1-\epsilon)c_{2})^{2} + y_{2}^{2} & = & \left | {\epsilon \over {2}}c_{1}
(1+\omega^{-1}_{l}) \right |^{2}
\end{eqnarray} 

where $\omega_{l}$ is defined as $\omega_l = e^{2 \pi i \over l}$ and $l : 1, 2
\ldots N$. 
 
Using the definition of  $\omega_l $ Eq. 43 reduces to 
\begin{eqnarray} 
 (x_{1}-(1-\epsilon)c_{1})^{2} + y_{1}^{2} = ({\epsilon}c_{2})^2 \cos^2( {\pi \over {l}}) \nonumber \\  
 (x_{2}-(1-\epsilon)c_{2})^{2} + y_{2}^{2} = ({\epsilon}c_{1})^2 \cos^2( {\pi \over {l}})   
\end{eqnarray} 

The radii on the r.h.s. of Eq. 44 have a maximum value for $l=1$ for which the equations
of the Gerschgorin circles
obtained are the same as those obtained in Eqs. 42. But $l=1$ is
just the basic $k=2$ period ( two lattice site case).
Thus the bounds on the largest eigenvalue obtained for a spatial period 
two solution for a lattice of $2N$ sites are the same as that obtained for
the spatial period two solution for a two lattice site case and it is
sufficient to look at the stability matrix of the basic spatial period ($k =2$)
case for the eigen-value analysis.

\section{Numerical Analysis}
We carry out the numerical stability analysis for spatiotemporal solutions with
the same spatiotemporal period i.e solutions of the kind 
spatial period 3 and corresponding temporal period 3 and spatial period 2
and temporal period 2 and so on. Analysis is carried out
in terms
of the new variables namely $a_{t}^{k}(i)$ and $b_{t}^{k}(i)$. 

We define the following vector notation,

\begin{equation} 
{\vec f_a}(\vec a, \vec b, \vec \Omega) \rightarrow
\big\{ f_{1a}(\vec a, \vec b, \vec \Omega), f_{2a}(\vec a,\vec b, \vec \Omega),
\ldots, f_{Na}(\vec a, \vec b, \vec \Omega)
     \big \}
 \end{equation}
 
where $ {\vec f_a}(\vec a, \vec b, \vec \Omega)$ denotes the evolution
of $a_{t}^{k}(i)$ as given by Eq.12. 
and
\begin{equation}
{\vec f_b}(\vec a, \vec b, \vec \Omega) \rightarrow
\big\{ f_{1b}(\vec a, \vec b, \vec \Omega), f_{2b}(\vec a, \vec b, \vec \Omega),
\ldots, f_{Nb}(\vec a, \vec b, \vec \Omega)
     \big \}
 \end{equation}

where $ {\vec f_b}(\vec a, \vec b, \vec \Omega)$ denotes the evolution
of $b_{t}^{k}(i)$ given by Eq.13. 
 
 For a lattice of $kN$ sites, each $i$ denotes the lattice index and 
each component of $\vec a$ and  $\vec b$ is defined
by Eqs.12 and 13 respectively. Each $a$ and $b$ is now a vector
  of the form
 
\begin{eqnarray} 
 (\vec a) \rightarrow
  \big\{ a_{t}^{k}(1), a_{t}^{k}(2),  \ldots, a_{t}^{k}(kN)  \big \} \nonumber \\
 (\vec b) \rightarrow
 \big\{ b_{t}^{k}(1), b_{t}^{k}(2),  \ldots, b_{t}^{k}(kN)  \big \}
  \end{eqnarray} 
 
 and the parameter, $\Omega$, also a vector, is represented as
\begin{equation} 
 (\vec \Omega) \rightarrow
\big\{ \Omega (1),  \Omega (2), \ldots,  \Omega (N)  \big \}
 \rightarrow \big\{\Omega, \Omega, \ldots, \Omega \big\}
\end{equation}
 
The  $\Omega$'s, could  
have different values at different  sites, but in this case have  the same value at
each site.
For a 1-d array of coupled sine circle maps, 
the stability criterion is
obtained by examining    
the eigen values of the corresponding stability 
matrix. From Eq.20 we obtain the fact that the two blocks
in the linear stability matrix, the $M_{t}^{k}$ are
identical. We consider one of these in the analysis and
for the calculation of the eigenvalues. After simplification
we obtain a $k \times k$ matrix corresponding to
the $k$th spatial period and temporal period $Q$ given by
 \begin{equation}
 {S^{k}_{Q}} =  {\prod_{t = 1}^{Q}} \pmatrix{ {\partial f_{1a}} \over
  {\partial {a_{t}^{k}(1)}} &
 {\partial f_{1a}}  \over
 {\partial {a_{t}^{k}(2)}} & \cdots & {\partial f_{1a}}
 \over   
 {\partial {a_{t}^{k}{t}(k)}} \cr
 {\partial f_{2a}} \over
{\partial {a_{t}^{k}(1)}} &
{\partial f_{2a}}  \over    
{\partial {a_{t}^{k}(2)}} & \cdots & {\partial f_{2a}}
   \over 
{\partial {a_{t}^{k}(k)}} \cr
\vdots & \vdots & \vdots & \vdots \cr
 {\partial f_{ka}}  \over
 {\partial {a_{t}^{k}(1)}} &
 {\partial f_{ka}}  \over
  {\partial {a_{t}^{k}(2)}} & \cdots & {\partial f_{ka}}
   \over
  {\partial {a_{t}^{k}(k)}}\cr}
  \end{equation}

The second $M_{t}^{k}$ which is obtained is identical and
is obtained by calculating the entries $ {\partial f_{mb}}  \over
 {\partial {b_{t}^{k}(j)}}$  where $m : 1, 2 \ldots k$ and for each $m$, 
$j : 1,2 \ldots k$.

We consider any one of these matrices and then obtain the eigenvalues
of the simplified $M_{t}^{k}$.

   Let $\{ \lambda_{i} \}$ be the set of eigenvalues of the matrix $S_{t}$.
   For a period $Q$ orbit to be stable the eigenvalues of the
    matrix $S^{k}_{Q} < 1$.
 The largest eigenvalue crossing 1 defines
 the marginal stability condition.

  Thus for the higher order temporal periods,
  we seek the solution
   to the following set of equations
 \begin{eqnarray}
  {\vec f^Q}(\vec a, \vec b, \vec \Omega)  =   (\vec a) + {\vec P}\nonumber \\
  {\vec f^Q}(\vec a, \vec b, \vec \Omega)  =   (\vec b) + {\vec P}
   \end{eqnarray}

  as the conditions of closure, namely the differences $a_{t}^{k}(t)$
and the sums $b_{t}^{k}(i)$. Since we are looking for
spatiotemporally periodic solutions we check for spatial closure
too in the following way.
\begin{eqnarray} 
\vec f_{a}(i+k) = \vec f_{a}(i)  \nonumber \\ 
\vec f_{b}(i+k)  = \vec f_{b}(i)  
\end{eqnarray}
                   
The condition for marginal stability is given by 
   \begin{equation}
   \lambda^{largest}  =   1
   \end{equation}

We start with initial conditions corresponding to the particular spatiotemporal
period under study and obtain points in the $ \Omega - \epsilon - K$ phase
space where the given  spatiotemporal solution is closed and 
is stable. We find the stability  edges in the $\epsilon-\Omega$ plane for each value of $K$ using
bisection method upto an accuracy of $10^{-8}$. 

\subsection{The $\Omega - \epsilon - K$ plot}

We plot the regions of stability in 
the $\Omega - \epsilon - K$ phase space for spatial periods 
$k = 2, 3, 4$. We confine our search to  regions of stability for solutions which have the same spatial and temporal
period. The stability regions for such solutions form a set of Arnold tongues in the  $\Omega - \epsilon - K$ space 
(See Fig.3).
Similar tongues were seen earlier for the $k=1$ 
or synchronised solution for  
different temporal periods \cite{nc 96}. The tongues  for higher spatial
periods are contained within
the Arnold tongues of the spatially synchronized and
temporally periodic solution of the same temporal period. 
 The tongues are symmetric about $\Omega = 0.5$ as       
 in the case of the single circle map and the synchronised solution of the  
coupled sine circle map lattice.  

The study of two dimensional slices in the $\Omega- \epsilon$ space for a fixed
value of $K$ show that the $\epsilon$ tongue develops a finite width for values 
of $K$ lower than those at which the 
$\Omega $ tongue develops a finite width (See Fig.4(a)-(c)). 
Fig. 4(a) reveals that for a small value of $K=0.016$,  
the $\epsilon$ tongue  shows a width
while there is no
width in $\Omega$ yet. Figs.4(b) and 4(c) show slices
at $K=0.14$ and $K=0.15$ respectively illustrating the
opening of the $\Omega = {1 \over 2}$ tongue.
Figs.4(b) and 4(c) also show that when the $\Omega = {1 \over 2 }$ tongue 
opens, a finite width of $\epsilon$ at $\Omega=0.495$ and
$\Omega=0.505$ is seen.
We thus conclude that since the $\epsilon$ width is
observed for a lower value of $K$(= 0.015) than the $\Omega$
width($K=0.15$) spatial orbits stabilize faster than temporal
orbits. 

An interesting  new bifurcation is seen in the
$\Omega = {0 \over 1}$ and $ {1 \over 1 }$ tongue. 
Starting with spatial period two initial conditions, one would expect spatial period one, temporal period one or spatial
period two temporal period one solutions within this tongue \cite{nc
96}. Instead,  the tongue now also contains travelling wave solutions with spatiotemporal period two. 
They show widths in the $\Omega$ and $\epsilon$ space which
widen with increasing $K$ 
and arise as a result
of a spatiotemporal bifurcation of the
spatially synchronized and temporal period one solution.
We discuss these solutions in detail in a separate
section where we also obtain  the stability interval of the $\epsilon $ region
for $\Omega = 0$ and $K=1$ analytically.  

The tongues in the $\Omega-\epsilon-K$ phase plane are confined to $ \epsilon < 0.5$  i.e.
we obtain stable closed spatiotemporal solutions for $ \epsilon < 0.5 $.
We thus conclude that lower values of $ \epsilon $ or smaller couplings
lead to stable solutions where the spatial and temporal periods are equal.
However, the $K=0$ or the coupled shift map case shows marginally stable orbits at
rational values of $\Omega$ for all
values of $\epsilon$. 

\subsection{The $\Omega - \epsilon - {P \over {Q}}$ plot}

We plot the stable regions for solutions with temporal winding number
$P \over Q$  in the $\Omega - \epsilon$
space at $K=1$ for spatial periods $k = Q$ in Fig.4(a). 
We thus obtain the widths
of various spatiotemporal solutions, $ {P \over Q} = {1 \over 2},
{1 \over 3}, {2 \over 3}, {1 \over 4}, {3 \over 4}, {1 \over 5}, {2 \over 5},
{3 \over 5}, {4 \over 5} $
 in the $ \Omega - \epsilon $ space at $K=1$.  
This forms a series of steps of finite $\Omega-\epsilon$ width for 
the observed values of $P \over Q$. This is similar to the Devil's
staircase seen in the case of the single circle map and the synchronised
solution \cite{nc 96,bak 84}. 
However
the width of the step for each $P \over Q$ is no longer
independent of $\epsilon $ as it was for the synchronized case
but depends on $\epsilon $. 
The staircase is symmetric  about $\Omega =0.5$.
Fig. 6 shows a two dimensional plot of the $\Omega - \epsilon$ space at
$K=1$ for various values of the winding number ${P \over Q}$.
The travelling wave solution shows the largest width in $\epsilon$
at $\Omega =0$ which decreases as $\Omega$ increases.

\section{The travelling wave solution}

We now discuss the travelling wave solution seen in the $\Omega = 0$ and
$\Omega = 1$ tongues.
 Starting with spatial period two initial
conditions in the $\Omega = 0$ and $\Omega =1$ tongues
we expect to find spatial period one or spatial period two
solutions with a temporal fixed point solution.
However we find a spatial period two solution
which also has temporal period two in the $\Omega =0$
and $\Omega =1$ tongue which  has the travelling wave structure
as seen in Fig. 1(b). This is the new bifurcation we obtain in the
$\Omega = 0$ and $\Omega = 1$ tongues.
Starting with
spatial period two initial conditions we find 
after a certain critical value of $\epsilon$, travelling wave solutions
of the period two kind are obtained in the $\Omega = 0$ and
$\Omega = 1$ tongue.
We analytically show that stable travelling wave 
solutions appear after a critical value of $\epsilon$ and
obtain the width of the $\epsilon$ tongue for $\Omega =0$ and $K=1$. 
The 
independent variables of the problem, namely the 
differences and the sums 
turn out to be convenient variables for this analysis.

Since we find  a bifurcation from
a spatially synchronised temporal period one solution to a spatial
period two temporal period two solution of the travelling wave type
we consider a lattice of $2N$ sites and construct nearest neighbour 
differences and sums defined as
\begin{eqnarray}
 \tilde a_{t} (i)= \theta_{t}(i) - \theta_{t}(i+1) \nonumber \\
 \tilde b_{t}(i)= \theta_{t}(i) + \theta_{t}(i+1)
 \end{eqnarray}

The evolution eqns. for $\tilde a_{t}(i)$ and $\tilde  b_{t}(i)$ are given by
Eqs. 12 and 13 (with $k$ = 1).

\begin{eqnarray}
\tilde a_{t+1}(i)& =& (1-\epsilon)\Big(\tilde a_{t}(i) - {K \over \pi} \sin( \pi \tilde a_{t}(i))
\cos( \pi \tilde b_{t}(i)) \Big) \nonumber \\
&+& {\epsilon \over2}\Big(\tilde a_{t}(i+1) - {K \over \pi} \sin( \pi \tilde a_{t}(i+1))
\cos( \pi\tilde  b_{t}(i+1)) \Big) \nonumber \\
&+& {\epsilon \over2}\big(\tilde a_{t}(i-1) - {K \over \pi} \sin( \pi \tilde a_{t}(i-1))
\cos( \pi \tilde b_{t}(i-1) \big) \nonumber \hspace{0.2 in} mod\hspace{0.05 in} 1\\
\end{eqnarray}
and
\begin{eqnarray}
\tilde b_{t+1}(i)& =& (1-\epsilon)\Big(\tilde b_{t}(i) - {K \over \pi} \sin( \pi \tilde b_{t}(i))
\cos( \pi \tilde a_{t}(i)) \Big) \nonumber \\
&+& {\epsilon \over2}\Big(\tilde b_{t}(i+1) - {K \over \pi} \sin( \pi \tilde b_{t}(i+1))
\cos( \pi \tilde a_{t}(i+1))\Big)  \nonumber \\
&+& {\epsilon \over2}\Big(\tilde b_{t}(i-1) - {K \over \pi} \sin( \pi \tilde b_{t}(i-1))
\cos( \pi \tilde a_{t}(i-1)) \Big) \nonumber  + 2 \Omega  \hspace{0.1 in}
mod\hspace{0.05 in} 1  \\
\end{eqnarray}

Consider Fig. 1(b) which shows the travelling wave solution
for which the spatial and temporal period is two. This solution can be expressed in a very compact
form in terms of the sum and difference variables. The differences
$\tilde a_{t}(i)$'s are equal but no longer zero as in the
synchronized case. Again, the sums $\tilde b_{t}(i)$'s are equal 
to some constant as in the synchronized case, but in addition it is clear that $\tilde b_{t+1}(i)
= \tilde b_{t}(i)$ reflecting the travelling wave nature of the solution. 
 Thus,
\begin{eqnarray}
 \tilde a_{t}(i) = - \tilde a_{t}(i+1) \neq 0 \nonumber \\
 \tilde a_{t+1}(i) = \tilde a_{t}(i+1) = - \tilde a_{t}(i)
\end{eqnarray}
and
\begin{eqnarray}
 \tilde b_{t}(i) = \tilde b_{t}(i+1) \nonumber \\
 \tilde b_{t+1}(i) = \tilde b_{t}(i)=constant
\end{eqnarray}

The two conditions above are specific to the spatiotemporal period two solution of 
the travelling wave type. 
Similar equations can be set up 
 in general for 
travelling wave solutions of spatiotemporal period $k$.

We use evolution equations for $\tilde a_{t}(i)$ and
$\tilde b_{t}(i)$ and carry out a linear stability 
analysis about the travelling wave solutions which
are expressed in terms of Eqs. 56 and 57.  We obtain the stability
matrix by expanding about the solutions $\tilde a_{t} = constant$
and $\tilde b_{t} = constant$. The linear stability matrix is given by
\begin{equation}
J^{4N}_{t} =  \pmatrix{ \tilde A_{t}^{2N} & \tilde B_{t}^{2N}\cr
   \tilde B_{t}^{2N} & \tilde A _{t}^{2N}\cr}
   \end{equation}

where $ \tilde A_{t}^{2N}$ and $ \tilde B_{t}^{2N}$ are
$2N \times 2N$ matrices which are given by
\begin{equation}
 \tilde A_{t}^{2N} = \pmatrix{ \epsilon_{s}\tilde A_{t}(1) &
 \epsilon_{n} \tilde A_{t}(2) & 0 &
 \cdots & 0 & \epsilon_{n} \tilde A_{t}(2) \cr
{\epsilon_n}\tilde A_{t}(1) &
 {\epsilon_s}\tilde A_{t}(2)  &
{\epsilon_n}\tilde A_{t}(1) &
    0  & \cdots & 0 \cr
    0 &  
  \epsilon_{n}\tilde A_{t}(2) &
  \epsilon_{s}\tilde A_{t}(1) &
 \epsilon_{n}\tilde A_{t}(2) & 0 & \cdots \cr
 \vdots& \vdots& \vdots& \vdots& \vdots& \vdots  \cr
      \epsilon_{n}\tilde A_{t}(1) & 0 &
         \cdots & 0 &
      \epsilon_{n}\tilde A_{t}(1) &
  {\epsilon_s}\tilde A_{t}(2) \cr}
 \end{equation}
 and
\begin{equation}
 \tilde B_{t}^{2N} =  \pmatrix{ \epsilon_{s}\tilde B_{t}(1) &
 \epsilon_n \tilde B_{t}(2) & 0 &
 \cdots &
 0 & {\epsilon_n}\tilde B_{t}(2) \cr
 {\epsilon_n}\tilde B_{t}(1) &
 {\epsilon_s}\tilde B_{t}(2) &
 {\epsilon_n}\tilde B_{t}(1) & 0 &
   \cdots & 0 \cr
 0 & {\epsilon_n}\tilde B_{t}(2)
 & {\epsilon_s}\tilde B_{t}(1) &
{\epsilon_n}\tilde B_{t}(2) &
  0 & \cdots \cr
\vdots& \vdots& \vdots& \vdots& \vdots& \vdots \cr
  {\epsilon_n}\tilde B_{t}(1) & 0 & \cdots & 0 &
   {\epsilon_n}\tilde B_{t}(1) &
   {\epsilon_s}\tilde B_{t}(2) \cr}
\end{equation}
      where
     $\epsilon_s , \epsilon_{n}$ are given by Eq.17

Here, each
 \begin{eqnarray}
 \tilde A_{t}(i) = {\Big (} 1 - {K }\cos( \pi \tilde a_{t}(i))
  \cos( \pi \tilde b_{t}(i)) {\Big )}
  \end{eqnarray}
  and
 \begin{eqnarray}
 \tilde B_{t}(i)={K }\sin( \pi \tilde a_{t}(i)) \sin( \pi \tilde b_{t}(i))
\end{eqnarray}
      
Using the conditions $\tilde b_{t}(1)=\tilde b_{t}(2)$ and
  $\tilde a_{t}(1) = - \tilde a_{t}(2)$ (seen from definition
of the differences and Fig.1b) $ \tilde A ^{\prime }_{t}$ and
$ \tilde B ^{\prime }_{t}$ reduce to matrices given by  

\begin{equation}
 \tilde A^{{2N}}_{t} = \pmatrix{ \epsilon_{s}\tilde A_{t}^{\prime}(1) &
 \epsilon_{n} \tilde A_{t}^{\prime}(1) & 0 &
 \cdots  & 0 & \epsilon_{n} \tilde A_{t}^{\prime}(1) \cr
  \epsilon_{n}\tilde A_{t}^{\prime}(1) &
 \epsilon_{s} \tilde A_{t}^{\prime}(1) &  
\epsilon_{n}\tilde A_{t}^{\prime}(1) & 0 & \cdots & 0 \cr
0 &\epsilon_{n} \tilde A_{t}^{\prime}(1) &  
\epsilon_{s} \tilde A_{t}^{\prime}(1) & 
\epsilon_{n} \tilde A_{t}^{\prime}(1) &
 0 & \cdots \cr
  \vdots& \vdots& \vdots& \vdots & \vdots & \vdots \cr
\epsilon_{n}\tilde A_{t}^{\prime}(1) & 0 & \cdots & 0 &
\epsilon_{n}\tilde A_{t}^{\prime}(1) & 
 \epsilon_{s} \tilde A_{t}^{\prime}(1)\cr}
    \end{equation}
and
   
\begin{equation} 
 \tilde B^{{2N}}_{t} = \pmatrix{ \epsilon_{s}\tilde B_{t}^{\prime}(1) &
 - \epsilon_{n} \tilde B_{t}^{\prime}(1) & 0 & 
 \cdots & 0  & - \epsilon_{n} \tilde B_{t}^{\prime}(1) \cr   
   \epsilon_{n}\tilde B_{t}^{\prime}(1) &
 - \epsilon_{s} \tilde B_{t}^{\prime}(1) &
   \epsilon_{n}\tilde B_{t}^{\prime}(1) & 0 & 
\cdots & 0 \cr  
0 & - \epsilon_{n} \tilde B_{t}^{\prime}(1) & \epsilon_{s} \tilde B_{t}^{\prime}(1) & - \epsilon_{n} \tilde B_{t}^{\prime}(1)  & 0 & \cdots \cr
  \vdots& \vdots& \vdots& \vdots & \vdots & \vdots \cr
\epsilon_{n}\tilde B_{t}^{\prime}(1) & 0 & \cdots&
 0 & \epsilon_{n}\tilde B_{t}^{\prime}(1) & - \epsilon_{s} \tilde B_{t}^{\prime}(1)\cr}
    \end{equation}
and each
\begin{eqnarray}
 \tilde A^{\prime}_{t}(i) & = & {\Big (}1 - {K } \cos \pi \tilde a_{t}(1) \cos \pi \tilde b_{t}(1){\Big )} \nonumber \\
  \tilde B^{\prime}_{t}(i) & = & {K } \sin \pi \tilde a_{t}(1) \sin \pi \tilde b_{t}(1) \\ 
\end{eqnarray}
 Using simple matrix algebra,  $J_{t}^{4N}$ can be put into
a block diagonal form given by
\begin{equation}
J^{4N}_{t} = \pmatrix{ M_{t}^{2N}(+)& 0 \cr
    0& M_{t}^{2N}(-)\cr}
        \end{equation}
where 
\begin{eqnarray}
M_{t}^{2N} (+) = \tilde A_{t}^{2N} + \tilde B_{t}^{2N} \nonumber \\
M_{t}^{2N} (-) = \tilde A_{t}^{2N} - \tilde B_{t}^{2N} 
\end{eqnarray}

The matrices $M_{t}^{2N}(+)$ and $M_{t}^{2N}(-)$ are similar
( $M_{t}^{2N}(-)$ = $\pi$ $M_{t}^{2N}(+)$ $\pi$, where $\pi$ is
the permutation matrix) and thus have the
same characteristic polynomial and it is sufficient to consider
the eigenvalues of one of them.

Following the treatment outlined in Section II B for a
lattice of $2N$, we use a similarity transformation which is a direct product
of Fourier matrices of size $N \times N$ and $2 \times 2$ which reduces $M_{t}^{2N}(+)$ to a matrix of $N$ blocks, each block of size $2 \times 2$. For the travelling wave solution of spatial and temporal period two  $M_{t}^2(l)$ is 
given by

\begin{equation} 
M^{2}_{t}(l) = \pmatrix{ (1 - \epsilon)( \tilde A_{t}^{\prime}(1)+ \tilde B_{t}^{\prime}(1)) & {\epsilon \over {2}} (1+\omega_{l}) 
(\tilde A_{t}^{\prime}(1) - \tilde B_{t}^{\prime}(1)) \cr 
{\epsilon \over {2}}(1+\omega_{l}^{-1}) (\tilde A_{t}^{\prime}(1) + \tilde B_{t}^{\prime}(1)) & (1- \epsilon) (\tilde A_{t}^{\prime}(1) - \tilde B_{t}^{\prime}(1)) \cr} 
\end{equation} 
where $\omega_{l}$ is as defined before and $l:1, 2 \ldots N$.  
As also discussed in Section II B, this can be reduced to the analysis
of just two lattice sites and $M_{t}^2(1)$ is given by

\begin{equation} 
M^{2}_{t}(1) = \pmatrix{ (1 - \epsilon)( \tilde A_{t}^{\prime}(1)+ \tilde B_{t}^{\prime}(1)) & {\epsilon } (\tilde A_{t}^{\prime}(1) - \tilde B_{t}^{\prime}(1)) \cr 
{\epsilon }(\tilde A_{t}^{\prime}(1) + \tilde B_{t}^{\prime}(1)) & (1- \epsilon) (\tilde A_{t}^{\prime}(1) - \tilde B_{t}^{\prime}(1)) \cr}   
\end{equation}

The eigenvalues of this matrix are given by 
 
\begin{equation}
\lambda = {(1-\epsilon) \tilde A ^{\prime }_{t}(1)} \pm  \sqrt{{\epsilon}^{2}
{\tilde A ^{\prime }_{t}(1)}^{2} + (1- 2 \epsilon){\tilde B ^{\prime }_{t}(1)}^{2}} 
\end{equation}

The largest eigenvalue $\tilde \lambda$ is the one corresponding to the
 + sign and is given by
\begin{equation}
\tilde \lambda = {(1-\epsilon) (1 - K  \cos \pi \tilde a_{t}(1) \cos \pi \tilde b_{t}(1) )} + \sqrt{{\epsilon}^{2}
 {(1 - K \cos \pi \tilde a_{t}(1) \cos \pi \tilde b_{t}(1))}^{2} + (1- 2 \epsilon){(K \sin \pi \tilde a_{t}(1) \sin \pi \tilde b_{t}(1))}^{2}}
\label{etw}
\end{equation}

Using the condition for closure and the largest eigenvalue we can obtain
the widths of the $\epsilon$ interval for which stable travelling wave solutions are obtained.

\subsection{Illustration for the $\Omega=0$ case}

For $\Omega = 0$ and $K=1$,  substituting the travelling wave conditions given by Eqs. 56 and 57 in
the evolution equation for 
$\tilde b_{t}(i)$ namely Eqs. 54 and 55 we obtain
\begin{equation}
{1 \over \pi}\sin{\pi \tilde b_{t}(1)} \cos{\pi \tilde a_{t}(1)} = 0
\end{equation}
which implies  
\begin{equation}
\tilde b_{t}(1) = 0, 1,2, \ldots n\\
\end{equation}
or
\begin{equation}
\tilde a_{t}(1) = {1 \over 2}, {3 \over 2} \ldots {2n+1 \over 2} 
\end{equation}
 
If we consider Eq. \ref{etw} we observe that if $\tilde a_{t}(1)$ is $1 \over 2$
 ( $mod 1$)
and $\tilde b_{t}(1)$ arbitrary, the eigenvalue is $>1$ and thus unstable. Hence for stable travelling wave solutions at $\Omega = 0$ and
$K=1$ we consider
Eq. 74.

Using Eq. 74 in  Eq. 54, the evolution equation
for $\tilde a_{t}(i)$ we get, 
\begin{equation}
(1-\epsilon)\tilde a_{t}(1) + {(1-2\epsilon) \over {2 \pi}}\sin( \pi \tilde a_{t}(1)) =0
\end{equation}
We consider the case where $\tilde b_{t}(1) = 1$  
  and
$\tilde a_{t}(1)$ is arbitrary, a condition also observed in numerical
simulations. The case $b_t=0$ corresponds to the
smaller  eigenvalue and $b_t=2 $ and above are ruled out due to the fact that 
 $\theta_{t}(i)$ always lies between 0 and 1. Thus using 
$\tilde b_{t}(1) = 1$ and $ \tilde a_{t}(1)$ arbitrary, in Eq. 72,
 the
largest eigenvalue is given by
\begin{equation}
\tilde \lambda = 1+ \cos(\pi \tilde a_{t}(1))
\end{equation}

For the edge of the stable solution, we have 
\begin{equation} 
1+ \cos(\pi \tilde a_{t}(1)) = 1
\end{equation}
which gives 
\begin{equation}
\tilde a_{t}(1) = {1 \over 2}, {3 \over 2} \ldots {2n+1 \over
2}
\end{equation}
which is also what we obtained in Eq. 75.

Thus for a stable spatiotemporal period two travelling
wave solution, the edge is obtained for $\tilde a_{t}(1) = 0.5$ (mod 1). 
Using $\tilde a_{t}(1)=0.5$
 in Eq. 76 we obtain $\epsilon = {(\pi +1) \over {(2+ \pi)}}$ which gives 
$\epsilon = 0.805523$. We have also obtained these solutions numerically and 
Fig. 6
shows the travelling wave bifurcation along with other spatiotemporally
periodic solutions. Fig. 6 is a two dimensional
plot of the widths in $\epsilon$ for various values of $P \over Q$ at $K=1$.
Comparing the value obtained numerically for the lower edge of the interval
for $\Omega =0$ and $K=1$ we obtain a close agreement with computational
accuracy.
The travelling wave solution is discontinuous at $\epsilon = 1$ as at this value  we obtain
$\tilde a_{t}(i)=0$ which is a  synchronised solution and is thus no longer
a travelling wave solution.

Similarly the evolution equations
for $\tilde a_{t}(i)$ and $\tilde b_{t}(i)$ alongwith Eqs. 54,55 and 72
can be solved 
for  fixed $\Omega$ and $K$ 
to obtain the corresponding width in $\epsilon$.

\section{Conclusions and Discussion}

We have studied the stability of spatially periodic solutions in a
lattice of coupled sine circle maps. The stability analysis of 
a system of $kN$ lattice sites with $N$ copies of the basic spatial period
$k$ has been reduced to the analysis of $N=1$ copies of the basic
spatial period $k$ by an appropriate choice of variables, a similarity
transformation via  Fourier matrices and the use of the Gerschgorin
theorem.  
We have provided an explicit illustration of this for a spatial period
$2$ case and a $2N$ site lattice.
Our analysis is completely general and can be applied to other coupled
map lattice systems as well.

We have also studied stability of solutions where the spatial and
temporal periods are equal, and found their regions of stability in the
$K-\epsilon-\Omega$ space numerically. We find that the stability regions of such
solutions show an Arnold tongue structure in the parameter space. 
We also see a Devil's staircase of winding numbers at $K=1$ 
in the  $\Omega - \epsilon - {P \over Q}$ space. A new and interesting
feature observed in the case of the $0 \over 1$ and $1 \over 1$ tongues
is the existence of a bifurcation to a spatial period two, temporal 
period two solution of the travelling wave type. The stability edge for
this solution can also be obtained analytically. The method can be
easily extended to travelling wave solutions in other coupled map
lattices. 

The framework set up by us  should be useful in the analysis of the
spatially periodic behaviour seen in a variety of situations  
like reaction-diffusion equations, coupled laser arrays, biological systems.
It should also be interesting  to investigate the correspondence between
the synchronised to travelling wave bifurcation seen by us and the in-phase to splay phase
bifurcation seen in Josephson Junction arrays \cite{weis 95}, anti-phase
behaviour in multimode lasers \cite{rroy 94}, or the synchronised to
rotating wave behaviour seen in rings of coupled chaotic oscillators
\cite{Matias}.   
We hope our study will find useful applications in some of these
contexts. 
 \vspace{0.2 in}

 {\bf ACKNOWLEDGMENTS}

\vspace{0.2 in}

We  thank the Institute of Mathematical Sciences, Madras,
 for computing
facilities.
One of the authors  (NC) gratefully acknowledges CSIR (India), for financial support.

\newpage
\begin{center}

\end{center}

\newpage

\begin{center}
{\bf FIGURE CAPTIONS}
\end{center}
\vspace{1.0 in}
\begin{center}
\begin{tabular}{lll}

Fig.1 & (a) & The spatially periodic solution with $k$ = 2 and $N$ = 3.\\
&&  $k$ is the basic spatial period, two in this case, which is
 repeated $N$ = 3 times ($kN$ = 6 sites) \\

&&\\

Fig.1 & (b) & The travelling wave solution with spatial period $k$ = 2 and
 and temporal period = 2.\\ && In this case the number of replicas
is $N$ = 3.\\ && It can be seen that $a_{t}(i) = - a_{t+1}(i) = a_{t}(i+1) $, $b_{t}(i) = b_{t+1}(i)
= b_{t}(i+1) $ \\

&&\\

Fig.2 && Illustration of Gerschgorin's theorem for the spatial period
two ($ k=2$) \\ && and temporal period one case.The eigenvalue is analytically
obtained to be 0.174334 and \\ && 0.193911. For $\epsilon =0.1 $ and 
$\Omega = 0$ and at $K=1$ we
obtain the stability matrix for $k$ (2)  sites \\ && and 
 construct circles
with radii 0.214 and 0.1929. in accordance with Eq.42. \\
&& The eigenvalue has been marked with an X  \\

&&\\

Fig.3 && Arnold tongues for spatiotemporally periodic solutions in a coupled
sine circle map lattice. \\ && The tongues have been plotted for 
winding numbers starting from ${ P \over Q}$
= $0 \over 1$,
 ${1 \over 1}$,${ 1 \over 2}$, ${1 \over 3}$, ${2 \over 3}$ as a function
of $\Omega$ and $\epsilon$.\\ 
&& We also observe
an additional new bifurcation within the tongue at $\Omega = 0$ and 
$\Omega = 1$.\\
&& This is a 
travelling wave solution (dashed lines) with spatial period two and temporal period two \\ && and has a structure as shown in Fig.1(b). \\

&&\\

Fig.4 & (a) & The opening of the $\epsilon$ tongue. At $K=0.016$ we observe
a finite width in \\ && $\epsilon$ at $\Omega = {1 \over 2} = 0.5$. The $\epsilon$
tongue opens before the $\Omega $ tongue \\ && which implies spatially periodic
solutions stabilize faster than temporally periodic solutions. \\

&&\\

Fig.4 & (b) & $\Omega = { 1 \over 2 } $ and $ \Omega = {1 \over 3 } $ at $K=0.14$ which show finite widths in $\epsilon$ but still no width in $\Omega$.\\

&&\\

Fig.4 & (c) &  $\Omega$ values as in Fig.4 (b), now with a finite width 
in $\Omega$ also for $K=0.15$.\\ 

&&\\

Fig.5 && The Devil's  staircase for spatiotemporally periodic solutions 
with the same winding number $ {P \over Q} $ \\ &&
in a coupled sine circle map lattice. The symmetry about
 $\Omega = 0.5$ is clearly seen.\\ && The widths are calculated
for
 $\Delta \Omega ({P \over Q})$ for ${ P \over Q}$ = ${0 \over 1}$,
 ${1 \over 1}$,${ 1 \over 2}$, ${1 \over 3}$, ${2 \over 3}$,
 ${1 \over 4}$, ${3 \over 4}$, ${1 \over 5}$, ${2 \over 5}$, ${3 \over 5}$,
 ${4 \over 5}$ \\ && We see an additional new width at
$ { P \over Q}$ = $0 \over 1$, ${1 \over 1}$. \\ &&  These are travelling
wave solutions (dashed lines) with spatial period two and temporal period two.\\ 

&&\\

Fig.6 && A two dimensional slice of the Devil's staircase for
the spatiotemporally periodic solutions \\
&&  with winding numbers as
in Fig.5. The shape and size of the steps is now dependent
on $\epsilon$ \\ && and symmetry about $\Omega = 0.5 $ is clearly seen. \\

\end{tabular}

\end{center}


\begin{thebibliography}{Dillo 83}
\bibitem{FN}We shall assign the superscript `d'
 to all stability matrices $J_{t}^{d}$ where $d$ denotes the dimension
of
the matrix and $J_{t}^{d}$ is a $d \times d$ matrix.
\end{thebibliography}
\end{document}